\documentclass[twocolumn]{aastex63}
\usepackage[utf8]{inputenc}
\usepackage{amsmath}
\usepackage{amssymb}
\usepackage{xcolor}
\usepackage{multirow}
\usepackage{soul}

\newcommand{\Emax}{E_\text{max}}
\newcommand{\Emaxl}{E_\text{max,l}}
\newcommand{\Emaxh}{E_\text{max,h}}
\newcommand{\alphal}{\alpha_\text{l}}
\newcommand{\alphah}{\alpha_\text{h}}
\newcommand{\Lc}{L_\text{c}}
\newcommand{\dtFlare}{\Delta t_\text{flare}}
\newcommand{\dtAgn}{\Delta t_\text{AGN}}
\newcommand{\dtIgmf}{\Delta t_\text{IGMF}}
\newcommand{\txs}{TXS~0506+056}

\received{XXX}
\revised{XXX}
\accepted{XXX}

\submitjournal{Astrophysical Journal Letters}

\shorttitle{Constraints on IGMFs from the TXS 0506+056 flare}
\shortauthors{Alves Batista, Saveliev}

\begin{document}

\title{Multimessenger Constraints on Intergalactic Magnetic Fields from the Flare of TXS 0506+056}

\email{r.batista@astro.ru.nl, andrey.saveliev@desy.de}

\author[0000-0003-2656-064X]{Rafael Alves Batista}
\affiliation{
Radboud University Nijmegen,
Department of Astrophysics/IMAPP,
6500 GL Nijmegen, The Netherlands}

\author[0000-0002-2734-6488]{Andrey Saveliev}
\affiliation{
Immanuel Kant Baltic Federal University,
Institute of Physics, Mathematics and Information Technology,
236041 Kaliningrad, Russia}
\affiliation{
Lomonosov Moscow State University,
Faculty of Computational Mathematics and Cybernetics,
119991 Moscow, Russia}

\begin{abstract}
The origin of magnetic fields in the universe is an open problem. Seed magnetic fields possibly produced in early times may have survived up to the present day close to their original form, providing an untapped window to the primeval universe. The recent observations of high-energy neutrinos from the blazar TXS~0506+056 in association with an electromagnetic counterpart in a broad range of wavelengths can be used to probe intergalactic magnetic fields via the time delay between the neutrinos and gamma rays as well as the time dependence of the gamma-ray fluxes. Using extensive three-dimensional Monte Carlo simulations, we present a novel method to constrain these fields. We apply it to TXS~0506+056 and, for the first time, derive constraints on both the magnetic-field strength and its coherence length, considering six orders of magnitude for each.
\end{abstract}


%
\section{Introduction} \label{sec:intro}

A long-standing problem in cosmology concerns the origin of magnetic fields in the universe. Two broad classes of mechanisms to explain magnetogenesis exist. Primordial (or cosmological) mechanisms posit that global processes taking place in the early universe could give rise to seed magnetic fields. Examples of such processes are inflation and phase transitions such as the electroweak and the quantum chromodynamics phase transitions (see~\citealt{durrer2013a} for a review). Astrophysical mechanisms, on the other hand, suggest that small-scale processes during the formation of structures gave rise to magnetic fields.

A way to distinguish between primordial and astrophysical mechanisms is to look for signs of magnetisation in cosmic voids. A negative signal would favour an astrophysical origin, but a positive one would not necessarily provide unambiguous evidence of a cosmological origin, since winds/outflows could carry magnetised material to the intergalactic medium (IGM). Nevertheless, the contamination of the IGM by winds and outflows is limited, and the seed magnetic fields far from structures, near the centre of voids, should remain in their pristine form~\citep{furlanetto2001a,bertone2006a}. Moreover, primordial and astrophysical mechanisms lead to distinct magnetic power spectra and hence different coherence lengths.

Faraday rotation measures of distant objects~\citep{vallee1991a} and the cosmic microwave background (CMB)~\citep{jedamzik2018a} provide the respective upper limits of $B \lesssim 10^{-9} \; {\rm G}$ and $B \lesssim 10^{-11} \; {\rm G}$ on the strength of intergalactic magnetic fields (IGMFs) at large scales. Gamma-ray-induced electromagnetic cascades yield, in general, the lower limit $B \gtrsim 10^{-16.5} \; {\rm G}$~\citep{neronov2010a,tavecchio2010a,taylor2011a,vovk2012a,fermi2018a}, with some additional exclusion regions between $10^{-15}$ and $10^{-13} \; \text{G}$ from the absence of extended emission around objects~\citep{hess2014a,veritas2017a}. The coherence length of IGMFs ($\Lc$) is poorly constrained, lying in the range $10^{-12} \lesssim L_c / {\rm Mpc} \lesssim 10^{3.5}$, the lower bound corresponding to the resistivity time scale of the universe, and the upper limit referring to the Hubble horizon.

The detection of the high-energy neutrino event IC-170922A by the IceCube Observatory~\citep{icecube2018a,icecube2018b} together with electromagnetic counterparts at various wavelengths~\citep{icecube2018b} marked the dawn of neutrino astronomy. These observations have been associated with a flare of the blazar \txs, located at redshift $z \approx 0.3365 \pm 0.0010$~\citep{paiano2018a}. The Major Atmospheric Gamma-ray Imaging Cherenkov (MAGIC) telescopes measured the flux at $E > 9 \times 10^{10} \; {\rm eV}$ indicating two periods of enhanced activity: MJD 58029.22 and MJD 58030.24~\citep{magic2018a}. The hypothesis that the correlation between the electromagnetic and the neutrino signals happened by chance is rejected at a  $3\sigma$-level~\citep{icecube2018b}. 

%
\section{Electromagnetic cascades and IGMFs}

High-energy gamma rays may initiate electromagnetic cascades in the intergalactic medium. \citet{aharonian1994a} and \citet{plaga1995a} suggested that they can be used to measure IGMFs. The underlying physics of this process is well known. A blazar emits high-energy gamma rays, which can interact with pervasive radiation fields such as the extragalactic background light (EBL) and the CMB, producing electron-positron pairs: $\gamma + \gamma_{\rm bg} \rightarrow e^+ + e^-$, with $\gamma_{\rm bg}$ denoting the background (CMB/EBL) photon. Electrons and positrons upscatter CMB/EBL photons to high energies via inverse Compton ($e^\pm + \gamma_{\rm bg} \rightarrow e^\pm + \gamma$). The high-energy photons produced will then restart the whole process, creating a cascade of particles. The cascade will stop when the energy of the photons drop below the kinematic threshold for pair production. The electrons and positrons are sensitive to the local magnetic field where they were produced. They are deflected in opposite directions, proportionally to the magnetic-field strength. 

Most of the observed high-energy gamma rays from cosmologically distant objects are attenuated and the spectrum measured between~GeV and~TeV energies is a combination of both prompt and cascade gamma rays. The former are produced at the source and do not interact during propagation, whereas the latter are secondaries from high-energy primaries that underwent a cascade process in the intergalactic medium. The spectrum of {\txs} indicates that at least a fraction of the flux at $\sim \,$GeV-TeV is comprised of secondary photons produced during propagation, as we will show later.
We do not expect a significant emission in TeV and above because the high-energy part of the spectrum is absorbed by the EBL, given the distance of the object. 
Note that the spectrum should extend up to energies of at least $\sim 400 \; \text{GeV}$, as observed by MAGIC. The High-Energy Stereoscopic System (H.E.S.S.) and the Very Energetic Radiation Imaging Telescope Array System (VERITAS) have also observed this object, but only upper limits were provided~\citep{icecube2018b}.

The flaring activity of {\txs} in a broad wavelength band started in 2017 June, lasted for about six months, and was preceded and succeeded by a quiescent period. The peak luminosity was reached around the time of detection of IC-170922A, decreasing slowly thereafter. It is tempting to directly correlate the very-high-energy gamma-ray signals observed by MAGIC with IC-170922A and to attempt to directly measure the strength of IGMFs as a function of the coherence length. However, multi-TeV gamma rays need not be produced simultaneously with the enhanced neutrino emission. In fact, they can be produced anytime during acceleration; nevertheless, the neutrino and TeV gamma-ray peaks lie within the same time window~\citep{gao2019a}. Therefore, we fix the duration of the flare: $\dtFlare \approx 6 \; {\rm months}$.

In the absence of IGMFs, both the neutrino and the electromagnetic signals would be detected roughly within the same time interval, $\dtFlare$. Any time delay incurred by IGMFs would be shorter than the period of flaring activity, otherwise the light curves of gamma rays would be considerably different than those at other wavelengths. Thus, the observation of very-high-energy gamma rays in coincidence with the flare sets an upper bound on the maximum time delay due to IGMFs: $\dtIgmf < \dtFlare$. This provides limits on the strength and coherence length of IGMFs.

%
\section{Model and simulations} 

Here we employ three-dimensional Monte Carlo simulations. The development of electromagnetic cascades in the intergalactic medium is modelled using the CRPropa~3 code~\citep{alvesbatista2016b} considering all relevant interactions and energy loss process, namely pair production, inverse Compton scattering, synchrotron emission, and adiabatic losses due to the expansion of the universe. The spectrum of gamma rays emitted by {\txs} is assumed to be a power law with spectral index $-\alpha$ and an exponential cut-off at $\Emax$, i.e., $E^{-\alpha} \exp(-E / \Emax)$. Several scenarios were studied, fixing $\Emax$ ($10^{10} \leq \Emax / {\rm eV} \leq 10^{14}$) and assuming $0 \leq \alpha \leq 4$. The following EBL models are used: \citet{gilmore2012a}, \citet{dominguez2011a}, and the upper and lower limits by ~\citet{stecker2016a}. For all cases studied, secondary photons produced in the cascade contribute to the observed flux at $\sim {\rm GeV}$ to at least a percent level. A number of scenarios for the magnetic field were considered, including all combinations of $10^{-19} \leq B / \text{G} \leq 10^{-14}$ and $10^{-2} \leq \Lc / \text{Mpc} \leq 10^{3}$, both in logarithmic steps of 1, in addition to the case $B=0$.

The neutrino emission takes place during a high state of the object. The spectral parameters of this period are not necessarily the same as the ones during the normal (low) state. Cascade photons stemming from gamma rays emitted during the low state may contribute to the observed flux at 
$E \gtrsim 1\,{\rm GeV}$, together with the flux from the high state. Thus, the spectrum of gamma rays effectively emitted by the source is:
\begin{equation} 
	\frac{dN}{dE} \propto  E^{-\alphal} \exp\left( -\frac{E}{\Emaxl} \right) + \eta E^{-\alphah} \exp\left( -\frac{E}{\Emaxh} \right),
	\label{eq:spec}
\end{equation}
wherein $\eta$ denotes the flux enhancement in the high state (subscript `h') with respect to the low state (`l'). This is computed within time interval $\dtFlare$, fixed by the duration of the neutrino flare. Another relevant parameter is the time scale over which {\txs} is a gamma-ray emitter at the low state, denoted by $\Delta t_{\rm AGN}$. Typically, AGN activity times range between~$10^6$ and~$10^8$ years~\citep{parma2002a}. We use $\Delta t_{\rm AGN} = 10$, $10^4$, and $10^7$ years. 

We estimated how much of the total flux comprises secondary photons produced in the cascade. To this end, we assumed $B=0$, and took only one of the components of equation~\ref{eq:spec}, which is equivalent to setting $\eta=0$. For the conservative case wherein $\Emax = 10^{11.5} \; \text{eV}$ and $\alpha < 3$, our simulation results suggest that at 10~GeV, at least 10\% of the flux correspond to cascade photons. This increases for stronger EBL models like the \citet{stecker2016a}, upper limit. Therefore, with the observations by MAGIC extending up to $E \sim 400 \; \text{GeV}$, we expect a sizeable contribution of secondary photons to the total flux, as shown by \citet{saveliev2020a}.

Photohadronic and hadronuclear interactions in blazar jets can produce neutrinos, as well as gamma rays of similar energies. According to most models the maximum cosmic-ray energy that can explain the neutrino observations is $E_{\rm CR} \sim 10^{16} \; {\rm eV}$~\citep{magic2018a,keivani2018a,murase2018a,liu2019a,gao2019a}. Consequently, neutrinos and gamma rays with $E \sim 10^{15} \; \text{eV}$ should be produced. This energy can be significantly degraded if the source environment is opaque to high-energy gamma rays. We do not concern ourselves with this absorption; instead, our phenomenological model describes a gamma-ray flux that is injected into the intergalactic medium \emph{after} escaping the object. 

\section{Fit results}

\begin{figure}
	\centering
	\includegraphics[width=.48\textwidth]{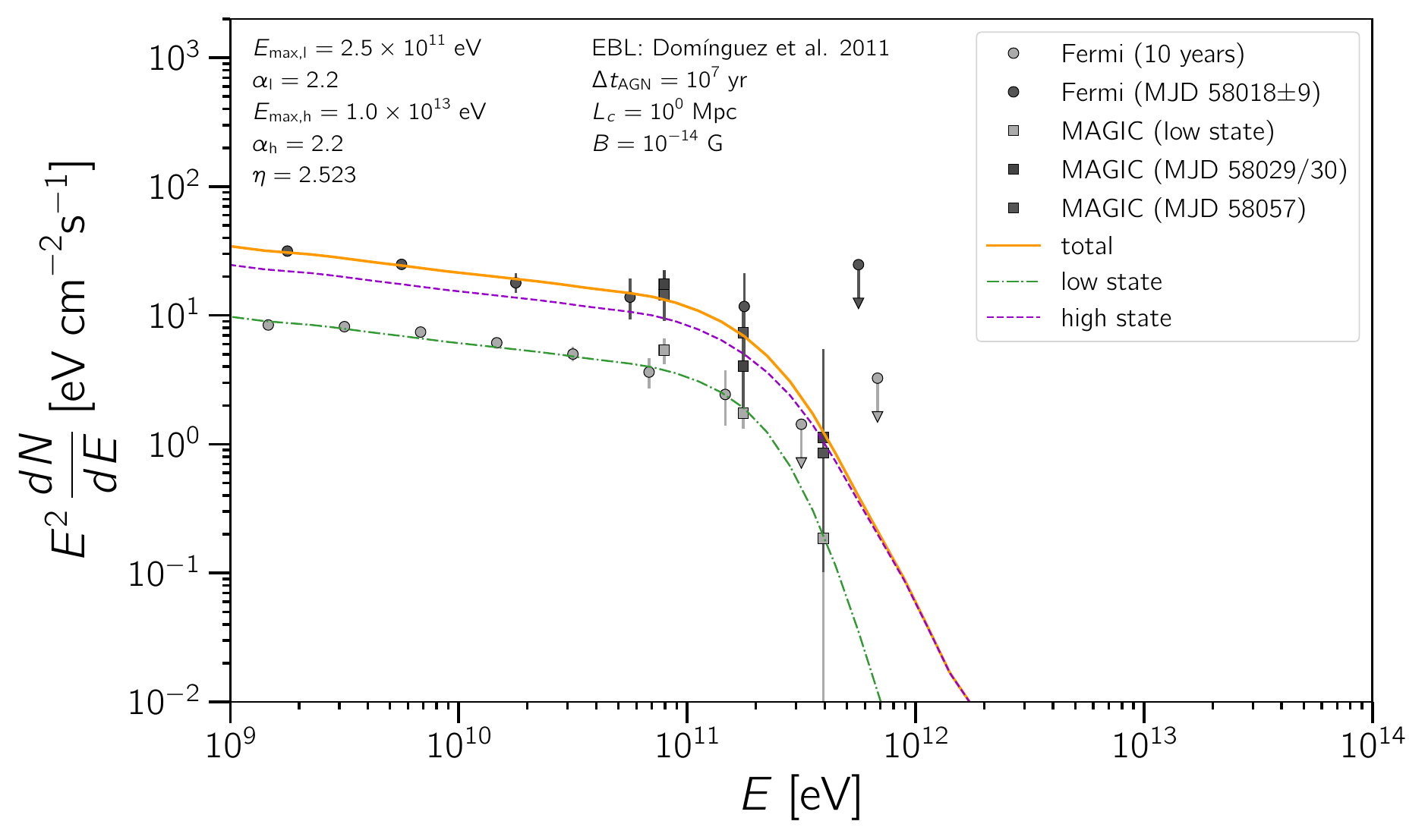}
	\caption{An example of the fit to the observed flux. The dot-dashed green line represents the fit to the low state. The dashed purple line corresponds to the high state ignoring the contribution of the low state. The combined total (low$+$high state) is shown as a thick orange solid line.}
	\label{fig:spec}
\end{figure}

We are now able to constrain the strength of IGMFs using information from both messengers, gamma rays and neutrinos. We first fit the spectrum for the low state. We find that, in this case, the spectral parameters remain virtually unaltered regardless of the magnetic-field properties: $\alphal=2.2$ and $\Emaxl = 250 \; \text{GeV}$ if we only consider combinations of $\alpha_\text{l}$ and $\Emaxl$ for which the fit produces P-values $p > 10^{-3}$. The second step is to use these values to scan the all the combinations of the parameters $\Emaxh$, $\alphah$, $B$, and $\Lc$. One example of the fitted spectrum is shown in figure~\ref{fig:spec}.

In the scope of this work we are interested in the case of a non-vanishing magnetic field, such that as a first step we carried out a hypothesis test for each of the four considered EBL models, with the null hypothesis being $B=0$. To do so, we marginalized over all other quantities, obtaining a probability distribution for the seven values of $B$ simulated. We find that only for the models by~\citet{dominguez2011a} and the lower limit model by~\citet{stecker2016a} the null hypothesis can be rejected. For completeness, we also included the EBL models by~\citet{gilmore2012a} and the upper limit by~\citet{stecker2016a} in our considerations, which, based on the likelihood ratio analysis we carried out, do not disfavour the $B=0$ case, but still allow for non-zero magnetic fields. This caveat should be borne in mind hereafter.

\begin{figure*}
	\centering
	\includegraphics[width=.48\textwidth]{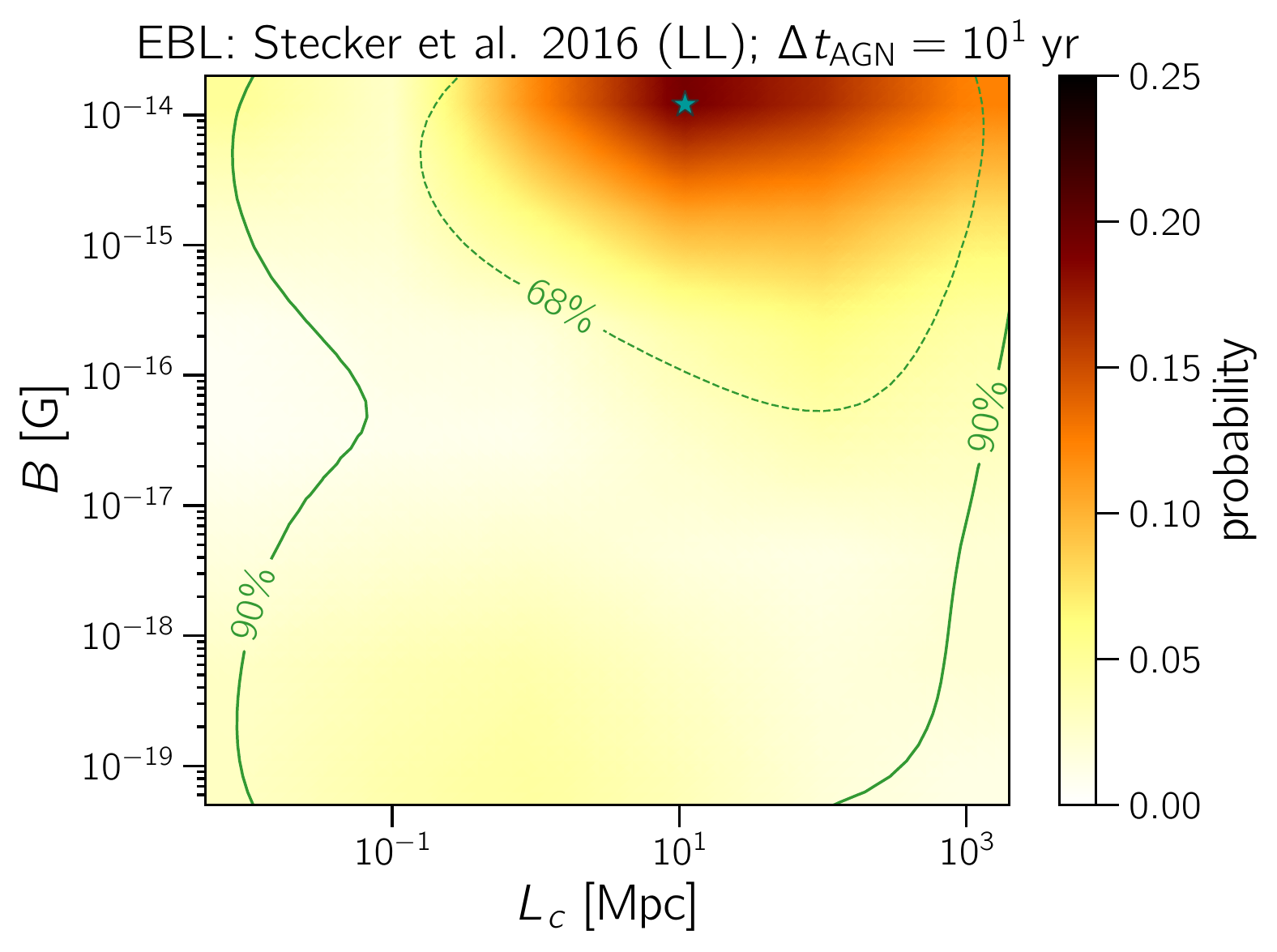}
	\includegraphics[width=.48\textwidth]{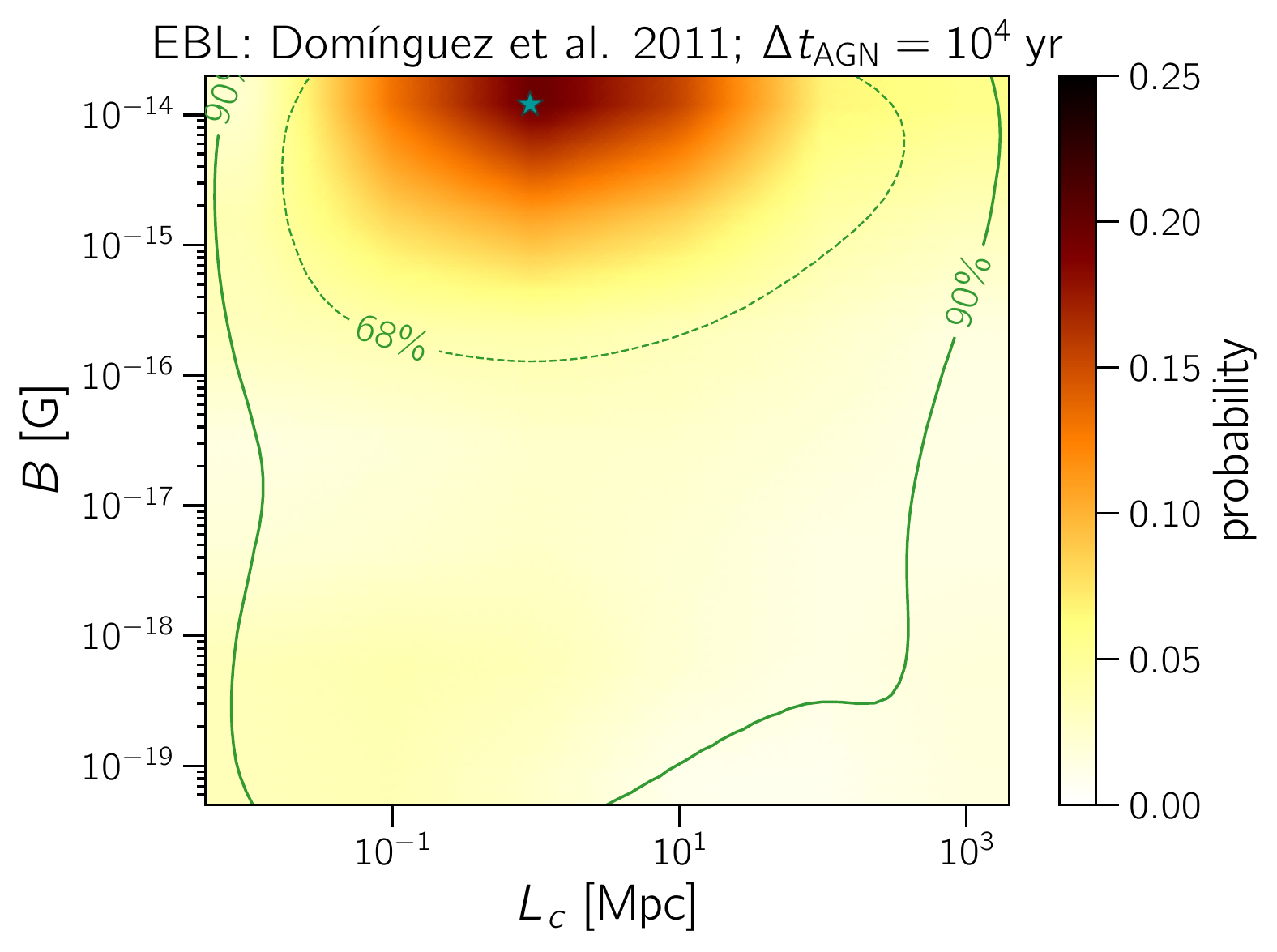}
	\caption{Results of the fit marginalised over the spectral parameters ($\Emax$ and $\alpha$) for the EBL models and $\dtAgn$ indicated in the figures. The colour scale denotes the probability normalised to unit. The star indicates the best-fit point.}
	\label{fig:fitB}
\end{figure*}

To constrain the magnetic field ($B$) and coherence length ($\Lc$), we first marginalise our results over the spectral parameters. Then we derive two-dimensional marginalised confidence regions for $B$ and $\Lc$, as shown in figure~\ref{fig:fitB}. The results for other values of $\dtAgn$ are very similar. In fact, we found this parameter to have little influence on the constraints. A summary with the best-fit intervals is shown in table~\ref{tab:fit}. Note, again, that not all models allow us to reject the null hypothesis ($B=0$).

\begin{table}
\centering
\caption{Best-fit parameters for all EBL models: \citet{stecker2016a} (S16l, S16u for lower and upper limits, respectively), \citet{gilmore2012a} (G12), and \citet{dominguez2011a} (D11).}
\begin{tabular}{cccc}
\hline 
\hline 
EBL & $\Delta t_\text{AGN} \; \text{[yr]}$  & $\log(L_\text{c}/\text{Mpc})$  & $\log(B/\text{G})$  \\ 
\hline 
S16l & $10^{1}$ & $1.0^{+1.3}_{-1.6}$ & $-15.3^{+1.0}_{-2.7}$ \\S16l & $10^{4}$ & $1.0^{+1.3}_{-1.6}$ & $-15.2^{+0.9}_{-2.7}$ \\S16l & $10^{7}$ & $1.0^{+1.3}_{-1.6}$ & $-15.2^{+0.9}_{-2.7}$ \\S16u & $10^{1}$ & $0.2^{+2.1}_{-1.6}$ & $-15.4^{+1.1}_{-2.5}$ \\S16u & $10^{4}$ & $0.0^{+2.4}_{-1.4}$ & $-15.2^{+0.9}_{-2.3}$ \\S16u & $10^{7}$ & $-0.1^{+2.3}_{-1.3}$ & $-15.2^{+0.9}_{-2.6}$ \\G12 & $10^{1}$ & $0.5^{+1.6}_{-1.7}$ & $-15.8^{+1.3}_{-2.4}$ \\G12 & $10^{4}$ & $0.6^{+1.6}_{-1.7}$ & $-15.6^{+1.2}_{-2.6}$ \\G12 & $10^{7}$ & $0.6^{+1.6}_{-1.7}$ & $-15.6^{+1.2}_{-2.6}$ \\D11 & $10^{1}$ & $0.2^{+1.5}_{-1.3}$ & $-15.4^{+1.0}_{-2.5}$ \\D11 & $10^{4}$ & $0.1^{+1.5}_{-1.3}$ & $-15.3^{+1.0}_{-2.6}$ \\D11 & $10^{7}$ & $0.1^{+1.5}_{-1.3}$ & $-15.3^{+1.0}_{-2.6}$ \\
\hline 
\hline 
\end{tabular}
\label{tab:fit}
\end{table}

Our results shown in figure~\ref{fig:fitB} provide seemingly weak constraints on the parameter space. For instance, with the EBL by~\citet{dominguez2011a}, the 90\% contour disfavours very large coherence lengths ($\Lc \gtrsim 100 \; \text{Mpc}$) for fields weaker than $B \sim 10^{-18} \; \text{G}$. If IGMFs have galactic scales ($\Lc \lesssim 10-100 \; \text{kpc}$), then $B \gtrsim 10^{-18} \; \text{G}$ (at 90\% C.L.). Similarly, for the~\citet{stecker2016a} lower-limit model, $\Lc \lesssim 10 \; \text{kpc}$ is not contained within the 95\% confidence region.

%
\section{Discussion}

IGMFs are commonly constrained using TeV-emitting extreme blazars (see e.g.~\citealt{neronov2010a}, \citealt{taylor2011a}, \citealt{arlen2014a}, \citealt{fermi2018a}). Due to EBL absorption, TeV fluxes from very distant objects are not observed, and only objects at redshifts $z \lesssim 0.20$ are used for this purpose. {\txs} is located at $z\approx 0.34$, so that any flux at TeV~energies would be suppressed. Consequently, the high-energy flux would be shifted to lower energies, retaining some information about the injection spectrum and the intervening magnetic fields. Alone, these constraints would likely be weak, but the picture changes with the temporal information provided by neutrinos.

Our results, shown in fig.~\ref{fig:fitB}, are compatible with bounds derived by other authors~\citep{neronov2010a,tavecchio2010a,taylor2011a,arlen2014a,fermi2018a}. This is not surprising given that we could not exclude most of the parameter space probed. Nevertheless, we successfully derived some constraints on the coherence length with our detailed three-dimensional treatment of gamma-ray propagation, whereas other works commonly rely in the small-angle approximation for calculating angular and temporal profiles (see e.g.~\citealt{fermi2018a}). Moreover, in these works the magnetic fields have a cell-like structure, whereas we consider a more realistic Kolmogorov turbulent field (details can be found in \citealt{alvesbatista2016a}). 

Similar analyses could be performed in a more clear fashion using GRBs~\citep{takahashi2008a,takahashi2012a}. In this case, there would be no low state and the number of free parameters in the fit would decrease. This analysis was carried out recently for GRB~190114C~\citep{wang2020a}. 

The~\citet{icecube2018a} reported activity near {\txs} in 2014/2015. An analysis of Fermi-LAT data does not provide any indications of enhanced activity around this period~\citep{fermi2019a}. Otherwise, we could have applied this same method to constrain IGMFs. 

Plasma instabilities may quench electromagnetic cascades propagating in the intergalactic medium~\citep{broderick2012a,sironi2014a,broderick2018a}. This is, however, under dispute (see e.g.~\citealt{miniati2013a}). Nevertheless, even if plasma instabilities are taken into account, this should not significantly affect our estimates because of the transient nature of the phenomenon, which would not allow the instability to grow fast enough over the brief period of enhanced activity~\citep{alvesbatista2019a}. 

Coherence lengths of $\sim 10 \; \text{kpc}$ are disfavoured at a 90\% confidence level for the two EBL models shown in fig.~\ref{fig:fitB}. This would disfavour models in which the IGM is magnetised by galactic winds (e.g.~\citet{bertone2006a}), since they predict $\Lc \sim 1-10 \; \text{kpc}$. Models in which IGMFs were generated by cosmic rays escaping galaxies prior to reionisation (e.g.,~\citet{miniati2011a}) are only marginally compatible with our results. Fields seeded by AGNs are expected to have $\sim$Mpc scales~\citep{durrer2013a}, and are well within the estimated bounds.
We have not probed $\Lc \lesssim 10 \; \text{kpc}$, a region of the parameter space that would allow us to constrain some cosmological magnetogenesis models~\citep{durrer2013a}.

The detection of high-energy neutrinos correlating with the electromagnetic signal favours hadronic models~\citep{magic2018a,gao2019a,keivani2018a} and strengthens the case for multi-TeV gamma-ray production in the blazar. The bounds presented here are rather robust and rely only on the assumption that the gamma rays at $E \gtrsim 1 \; {\rm GeV}$ observed by Fermi-LAT, and the highest-energy bin observed by MAGIC at $E \approx 400 \; {\rm GeV}$, are produced during the neutrino flaring period. Therefore, the reliability of our limits reflects the significance of this correlation. 

The multi-messenger approach used to constrain IGMFs based on time delays between high-energy gamma rays and their counterparts (in neutrinos and other wavelengths) is powerful. It differs from the methods commonly found in the literature~\citep{plaga1995a,neronov2009a,neronov2013a} in that we do not attempt to correlate the~GeV and~TeV emissions with each other. Instead, we use the time delay between the neutrino and the gamma-ray signals. For this reason, this method enables us to probe the universe up to high redshifts, since no~TeV signal is required to perform these estimates and we can evade the limitations placed by the EBL attenuation of the very-high-energy part of the spectrum.  

The same idea used here could be applied whenever a high-energy gamma-ray signal correlates with the arrival of high-energy neutrinos, since this guarantees the production of very energetic gamma rays \emph{in} the source, though there is no guarantee that they can escape the environment. In the absence of neutrinos, the time delays between the cascade photons and another messenger such as gravitational waves (e.g. from mergers of compact objects) would rely on the existence of a putative multi-TeV emission which, in the case of high-redshift sources, cannot be observed. Either way, gamma rays in the $\sim$GeV band, together with another messenger (neutrinos, gravitational waves, or $\sim$TeV gamma rays) can be used to place limits on \emph{both} the strength \emph{and} the coherence length of IGMFs.

%
\section{Summary and Outlook}

We have here derived combined constraints on \emph{both} the coherence length \emph{and} the strength of IGMFs using three-dimensional simulations. On one hand, the constraints we derived are relatively weak compared to the existing ones, given that we used only one object. On the other hand, this is the first time that bounds on the coherence length are obtained, which is of utmost importance for understanding the origin of IGMFs.

We showed that the intrinsic spectral parameters of the object are degenerate with respect to the magnetic-field ones. It follows that magnetic-field effects may be important when fitting the high-energy region of spectral energy distributions. We investigate this in more detail in another work~\citep{saveliev2020a}.

Our results exclude the hypothesis of null IGMFs for only two EBL models, out of the four tested. For these models, considering the range of parameters studied, we could obtain bounds on the coherence length. The upper bound was inferred to be $\Lc \lesssim 300 \; \text{Mpc}$. The lower bound depends on the EBL: $\Lc \gtrsim 30 \; \text{kpc}$ for the more general case and, more interestingly, $\Lc \gtrsim 300 \; \text{kpc}$ for a weak EBL, at a 90\% C.L.

\section*{Acknowledgements} 

RAB gratefully acknowledges the funding from the Radboud Excellence Initiative, and from the S\~ao Paulo Research Foundation (FAPESP) through grant \#2017/12828-4 in the early stages of this work. The work of AS was supported by the Russian Science Foundation under grant no. 19-71-10018, carried out at the Immanuel Kant Baltic Federal University. 
Part of the simulations were performed in the computing facilities of the GAPAE group at Institute of Astronomy, Geophysics and Atmospheric Sciences of the University of São Paulo, FAPESP grant: \#2013/10559-5.

\software{CRPropa~\citep{alvesbatista2016b}, SciPy~\citep{scipy2001a}, NumPy~\citep{vanderwalt2011a}, Matplotlib~\citep{hunter2007a}.}

\bibliography{references}

\begin{thebibliography}{}
\expandafter\ifx\csname natexlab\endcsname\relax\def\natexlab#1{#1}\fi
\providecommand{\url}[1]{\href{#1}{#1}}
\providecommand{\dodoi}[1]{doi:~\href{http://doi.org/#1}{\nolinkurl{#1}}}
\providecommand{\doeprint}[1]{\href{http://ascl.net/#1}{\nolinkurl{http://ascl.net/#1}}}
\providecommand{\doarXiv}[1]{\href{https://arxiv.org/abs/#1}{\nolinkurl{https://arxiv.org/abs/#1}}}

\bibitem[{{Ackermann} {et~al.}(2018){Ackermann}, {Ajello}, {Baldini}, {Ballet},
  {Barbiellini}, {Bastieri}, {Bellazzini}, {Bissaldi}, {Blandford}, {Bloom},
  {Bonino}, {Bottacini}, {Brandt}, {Bregeon}, {Bruel}, {Buehler}, {Cameron},
  {Caputo}, {Caraveo}, {Castro}, {Cavazzuti}, {Charles}, {Cheung}, {Chiaro},
  {Ciprini}, {Cohen-Tanugi}, {Costantin}, {Cutini}, {D'Ammando}, {de Palma},
  {Desai}, {Di Lalla}, {Di Mauro}, {Di Venere}, {Favuzzi}, {Finke},
  {Franckowiak}, {Fukazawa}, {Funk}, {Fusco}, {Gargano}, {Gasparrini},
  {Giglietto}, {Giordano}, {Giroletti}, {Green}, {Grenier}, {Guillemot},
  {Guiriec}, {Hays}, {Hewitt}, {Horan}, {J{\'o}hannesson}, {Kensei}, {Kuss},
  {Larsson}, {Latronico}, {Lemoine-Goumard}, {Li}, {Longo}, {Loparco},
  {Lovellette}, {Lubrano}, {Magill}, {Maldera}, {Manfreda}, {Mazziotta},
  {McEnery}, {Meyer}, {Mizuno}, {Monzani}, {Morselli}, {Moskalenko}, {Negro},
  {Nuss}, {Omodei}, {Orienti}, {Orlando}, {Ormes}, {Palatiello}, {Paliya},
  {Paneque}, {Perkins}, {Persic}, {Pesce- Rollins}, {Piron}, {Porter},
  {Principe}, {Rain{\`o}}, {Rando}, {Rani}, {Razzaque}, {Reimer}, {Reimer},
  {Reposeur}, {Sgr{\`o}}, {Siskind}, {Spandre}, {Spinelli}, {Suson}, {Tajima},
  {Thayer}, {Tibaldo}, {Torres}, {Tosti}, {Valverde}, {Venters}, {Vogel},
  {Wood}, {Wood}, {Zaharijas}, {Fermi-LAT Collaboration}, \&
  {Biteau}}]{fermi2018a}
{Ackermann}, M., {Ajello}, M., {Baldini}, L., {et~al.} 2018, The Astrophysical
  Journal Supplement Series, 237, 32, \dodoi{10.3847/1538-4365/aacdf7}

\bibitem[{{Aharonian} {et~al.}(1994){Aharonian}, {Coppi}, \&
  {Voelk}}]{aharonian1994a}
{Aharonian}, F.~A., {Coppi}, P.~S., \& {Voelk}, H.~J. 1994, The Astrophysical
  Journal Letters, 423, L5, \dodoi{10.1086/187222}

\bibitem[{{Alves Batista} {et~al.}(2019){Alves Batista}, {Saveliev}, \& {de
  Gouveia Dal Pino}}]{alvesbatista2019a}
{Alves Batista}, R., {Saveliev}, A., \& {de Gouveia Dal Pino}, E.~M. 2019,
  Monthly Notices of the Royal Astornomical Society, 489, 3836,
  \dodoi{10.1093/mnras/stz2389}

\bibitem[{{Alves Batista} {et~al.}(2016{\natexlab{a}}){Alves Batista},
  {Saveliev}, {Sigl}, \& {Vachaspati}}]{alvesbatista2016a}
{Alves Batista}, R., {Saveliev}, A., {Sigl}, G., \& {Vachaspati}, T.
  2016{\natexlab{a}}, Physical Review D, 94, 083005,
  \dodoi{10.1103/PhysRevD.94.083005}

\bibitem[{{Alves Batista} {et~al.}(2016{\natexlab{b}}){Alves Batista},
  {Dundovic}, {Erdmann}, {Kampert}, {Kuempel}, {M{\"u}ller}, {Sigl}, {van
  Vliet}, {Walz}, \& {Winchen}}]{alvesbatista2016b}
{Alves Batista}, R., {Dundovic}, A., {Erdmann}, M., {et~al.}
  2016{\natexlab{b}}, Journal of Cosmology and Astroparticle Physics, 5, 038,
  \dodoi{10.1088/1475-7516/2016/05/038}

\bibitem[{{Ansoldi} {et~al.}(2018){Ansoldi}, {Antonelli}, {Arcaro}, {Baack},
  {Babi{\'c}}, {Banerjee}, {Bangale}, {Barres de Almeida}, {Barrio}, {Becerra
  Gonz{\'a}lez}, {Bednarek}, {Bernardini}, {Berse}, {Berti}, {Besenrieder},
  {Bhattacharyya}, {Bigongiari}, {Biland}, {Blanch}, {Bonnoli}, {Carosi},
  {Ceribella}, {Chatterjee}, {Colak}, {Colin}, {Colombo}, {Contreras},
  {Cortina}, {Covino}, {Cumani}, {D'Elia}, {Da Vela}, {Dazzi}, {De Angelis},
  {De Lotto}, {Delfino}, {Delgado}, {Di Pierro}, {Dom{\'\i}nguez}, {Dominis
  Prester}, {Dorner}, {Doro}, {Einecke}, {Elsaesser}, {Fallah Ramazani},
  {Fattorini}, {Fern{\'a}ndez-Barral}, {Ferrara}, {Fidalgo}, {Foffano},
  {Fonseca}, {Font}, {Fruck}, {Gallozzi}, {Garc{\'\i}a L{\'o}pez},
  {Garczarczyk}, {Gaug}, {Giammaria}, {Godinovi{\'c}}, {Guberman}, {Hadasch},
  {Hahn}, {Hassan}, {Hayashida}, {Herrera}, {Hoang}, {Hrupec}, {Inoue},
  {Ishio}, {Iwamura}, {Konno}, {Kubo}, {Kushida}, {Lamastra}, {Lelas}, {Leone},
  {Lindfors}, {Lombardi}, {Longo}, {L{\'o}pez}, {Maggio}, {Majumdar},
  {Makariev}, {Maneva}, {Manganaro}, {Mannheim}, {Maraschi}, {Mariotti},
  {Mart{\'\i}nez}, {Masuda}, {Mazin}, {Mielke}, {Minev}, {Miranda}, {Mirzoyan},
  {Moralejo}, {Moreno}, {Moretti}, {Neustroev}, {Niedzwiecki}, {Nievas
  Rosillo}, {Nigro}, {Nilsson}, {Ninci}, {Nishijima}, {Noda}, {Nogu{\'e}s},
  {Paiano}, {Palacio}, {Paneque}, {Paoletti}, {Paredes}, {Pedaletti},
  {Pe{\~n}il}, {Peresano}, {Persic}, {Pfrang}, {Prada Moroni}, {Prandini},
  {Puljak}, {Garcia}, {Rhode}, {Rib{\'o}}, {Rico}, {Righi}, {Rugliancich},
  {Saha}, {Saito}, {Satalecka}, {Schweizer}, {Sitarek}, {{\v{S}}nidari{\'c}},
  {Sobczynska}, {Stamerra}, {Strzys}, {Suri{\'c}}, {Tavecchio}, {Temnikov},
  {Terzi{\'c}}, {Teshima}, {Torres-Alb{\'a}}, {Tsujimoto}, {Vanzo}, {Vazquez
  Acosta}, {Vovk}, {Ward}, {Will}, {Zari{\'c}}, \& {Cerruti}}]{magic2018a}
{Ansoldi}, S., {Antonelli}, L.~A., {Arcaro}, C., {et~al.} 2018, The
  Astrophysical Journal, 863, L10, \dodoi{10.3847/2041-8213/aad083}

\bibitem[{{Arlen} {et~al.}(2014){Arlen}, {Vassilev}, {Weisgarber}, {Wakely}, \&
  {Yusef Shafi}}]{arlen2014a}
{Arlen}, T.~C., {Vassilev}, V.~V., {Weisgarber}, T., {Wakely}, S.~P., \& {Yusef
  Shafi}, S. 2014, The Astrophysical Journal, 796, 18,
  \dodoi{10.1088/0004-637X/796/1/18}

\bibitem[{{Bertone} {et~al.}(2006){Bertone}, {Vogt}, \&
  {En{\ss}lin}}]{bertone2006a}
{Bertone}, S., {Vogt}, C., \& {En{\ss}lin}, T. 2006, \mnras, 370, 319,
  \dodoi{10.1111/j.1365-2966.2006.10474.x}

\bibitem[{Broderick {et~al.}(2012)Broderick, Chang, \&
  Pfrommer}]{broderick2012a}
Broderick, A.~E., Chang, P., \& Pfrommer, C. 2012, The Astrophysical Journal,
  752, 22, \dodoi{10.1088/0004-637X/752/1/22}

\bibitem[{{Broderick} {et~al.}(2018){Broderick}, {Tiede}, {Chang}, {Lamberts},
  {Pfrommer}, {Puchwein}, {Shalaby}, \& {Werhahn}}]{broderick2018a}
{Broderick}, A.~E., {Tiede}, P., {Chang}, P., {et~al.} 2018, The Astrophysical
  Journal, 868, 87, \dodoi{10.3847/1538-4357/aae5f2}

\bibitem[{{Dom{\'\i}nguez} {et~al.}(2011){Dom{\'\i}nguez}, {Primack},
  {Rosario}, {Prada}, {Gilmore}, {Faber}, {Koo}, {Somerville},
  {P{\'e}rez-Torres}, {P{\'e}rez-Gonz{\'a}lez}, {Huang}, {Davis},
  {Guhathakurta}, {Barmby}, {Conselice}, {Lozano}, {Newman}, \&
  {Cooper}}]{dominguez2011a}
{Dom{\'\i}nguez}, A., {Primack}, J.~R., {Rosario}, D.~J., {et~al.} 2011,
  Monthly Notices of the Royal Astronomical Society, 410, 2556,
  \dodoi{10.1111/j.1365-2966.2010.17631.x}

\bibitem[{{Durrer} \& {Neronov}(2013)}]{durrer2013a}
{Durrer}, R., \& {Neronov}, A. 2013, The Astronomy and Astrophysics Review, 21,
  62, \dodoi{10.1007/s00159-013-0062-7}

\bibitem[{{Furlanetto} \& {Loeb}(2001)}]{furlanetto2001a}
{Furlanetto}, S.~R., \& {Loeb}, A. 2001, \apj, 556, 619, \dodoi{10.1086/321630}

\bibitem[{{Gao} {et~al.}(2019){Gao}, {Fedynitch}, {Winter}, \&
  {Pohl}}]{gao2019a}
{Gao}, S., {Fedynitch}, A., {Winter}, W., \& {Pohl}, M. 2019, Nature Astronomy,
  3, 88, \dodoi{10.1038/s41550-018-0610-1}

\bibitem[{{Garrappa} {et~al.}(2019){Garrappa}, {Buson}, {Franckowiak},
  {Fermi-LAT Collaboration}, {Shappee}, {Beacom}, {Dong}, {Holoien},
  {Kochanek}, {Prieto}, {Stanek}, {Thompson}, {ASAS-SN Collaboration},
  {Aartsen}, {Ackermann}, {Adams}, {Aguilar}, {Ahlers}, {Ahrens}, {Alispach},
  {Andeen}, {Anderson}, {Ansseau}, {Anton}, {Arg{\"u}elles}, {Auffenberg},
  {Axani}, {Backes}, {Bagherpour}, {Bai}, {Barbano}, {Barwick}, {Baum}, {Bay},
  {Beatty}, {Becker}, {Becker Tjus}, {BenZvi}, {Berley}, {Bernardini},
  {Besson}, {Binder}, {Bindig}, {Blaufuss}, {Blot}, {Bohm}, {B{\"o}rner},
  {B{\"o}ser}, {Botner}, {Bourbeau}, {Bourbeau}, {Bradascio}, {Braun}, {Bretz},
  {Bron}, {Brostean-Kaiser}, {Burgman}, {Busse}, {Carver}, {Chen}, {Cheung},
  {Chirkin}, {Clark}, {Classen}, {Collin}, {Conrad}, {Coppin}, {Correa},
  {Cowen}, {Cross}, {Dave}, {de Andr{\'e}}, {De Clercq}, {DeLaunay},
  {Dembinski}, {Deoskar}, {De Ridder}, {Desiati}, {de Vries}, {de Wasseige},
  {de With}, {DeYoung}, {Diaz}, {D{\'\i}az-V{\'e}lez}, {Dujmovic}, {Dunkman},
  {Dvorak}, {Eberhardt}, {Ehrhardt}, {Eller}, {Evenson}, {Fahey}, {Fazely},
  {Felde}, {Filimonov}, {Finley}, {Franckowiak}, {Friedman}, {Fritz},
  {Gaisser}, {Gallagher}, {Ganster}, {Garrappa}, {Gerhardt}, {Ghorbani},
  {Glauch}, {Gl{\"u}senkamp}, {Goldschmidt}, {Gonzalez}, {Grant}, {Griffith},
  {G{\"u}nder}, {G{\"u}nd{\"u}z}, {Haack}, {Hallgren}, {Halve}, {Halzen},
  {Hanson}, {Hebecker}, {Heereman}, {Helbing}, {Hellauer}, {Henningsen},
  {Hickford}, {Hignight}, {Hill}, {Hoffman}, {Hoffmann}, {Hoinka},
  {Hokanson-Fasig}, {Hoshina}, {Huang}, {Huber}, {Hultqvist}, {H{\"u}nnefeld},
  {Hussain}, {In}, {Iovine}, {Ishihara}, {Jacobi}, {Japaridze}, {Jeong},
  {Jero}, {Jones}, {Kang}, {Kappes}, {Kappesser}, {Karg}, {Karl}, {Karle},
  {Katz}, {Kauer}, {Keivani}, {Kelley}, {Kheirandish}, {Kim}, {Kintscher},
  {Kiryluk}, {Kittler}, {Klein}, {Koirala}, {Kolanoski}, {K{\"o}pke}, {Kopper},
  {Kopper}, {Koskinen}, {Kowalski}, {Krings}, {Kr{\"u}ckl}, {Kulacz}, {Kunwar},
  {Kurahashi}, {Kyriacou}, {Labare}, {Lanfranchi}, {Larson}, {Lauber}, {Lazar},
  {Leonard}, {Leuermann}, {Liu}, {Lohfink}, {Lozano Mariscal}, {Lu},
  {Lucarelli}, {L{\"u}nemann}, {Luszczak}, {Madsen}, {Maggi}, {Mahn}, {Makino},
  {Mallot}, {Mancina}, {Mari{\textcommabelow s}}, {Maruyama}, {Mase}, {Maunu},
  {Meagher}, {Medici}, {Medina}, {Meier}, {Meighen-Berger}, {Menne}, {Merino},
  {Meures}, {Miarecki}, {Micallef}, {Moment{\'e}}, {Montaruli}, {Moore},
  {Moulai}, {Nagai}, {Nahnhauer}, {Nakarmi}, {Naumann}, {Neer}, {Niederhausen},
  {Nowicki}, {Nygren}, {Obertacke Pollmann}, {Olivas}, {O'Murchadha},
  {O'Sullivan}, {Palczewski}, {Pandya}, {Pankova}, {Park}, {Peiffer},
  {P{\'e}rez de los Heros}, {Pieloth}, {Pinat}, {Pizzuto}, {Plum}, {Price},
  {Przybylski}, {Raab}, {Raissi}, {Rameez}, {Rauch}, {Rawlins}, {Rea},
  {Reimann}, {Relethford}, {Renzi}, {Resconi}, {Rhode}, {Richman}, {Robertson},
  {Rongen}, {Rott}, {Ruhe}, {Ryckbosch}, {Rysewyk}, {Safa}, {Sanchez Herrera},
  {Sandrock}, {Sandroos}, {Santander}, {Sarkar}, {Sarkar}, {Satalecka},
  {Schaufel}, {Schlunder}, {Schmidt}, {Schneider}, {Schneider}, {Schumacher},
  {Sclafani}, {Seckel}, {Seunarine}, {Silva}, {Snihur}, {Soedingrekso},
  {Soldin}, {Song}, {Spiczak}, {Spiering}, {Stachurska}, {Stamatikos},
  {Stanev}, {Stasik}, {Stein}, {Stettner}, {Steuer}, {Stezelberger},
  {Stokstad}, {St{\"o}{\ss}l}, {Strotjohann}, {Stuttard}, {Sullivan},
  {Sutherland}, {Taboada}, {Tenholt}, {Ter-Antonyan}, {Terliuk}, {Tilav},
  {Tomankova}, {T{\"o}nnis}, {Toscano}, {Tosi}, {Tselengidou}, {Tung},
  {Turcati}, {Turcotte}, {Turley}, {Ty}, {Unger}, {Unland Elorrieta}, {Usner},
  {Vandenbroucke}, {Van Driessche}, {van Eijk}, {van Eijndhoven}, {Vanheule},
  {van Santen}, {Vraeghe}, {Walck}, {Wallace}, {Wallraff}, {Wandkowsky},
  {Watson}, {Weaver}, {Weiss}, {Weldert}, {Wendt}, {Werthebach}, {Westerhoff},
  {Whelan}, {Whitehorn}, {Wiebe}, {Wiebusch}, {Wille}, {Williams}, {Wills},
  {Wolf}, {Wood}, {Wood}, {Woschnagg}, {Wrede}, {Xu}, {Xu}, {Xu}, {Yanez},
  {Yodh}, {Yoshida}, {Yuan}, \& {IceCube Collaboration}}]{fermi2019a}
{Garrappa}, S., {Buson}, S., {Franckowiak}, A., {et~al.} 2019, The
  Astrophysical Journal, 880, 103, \dodoi{10.3847/1538-4357/ab2ada}

\bibitem[{{Gilmore} {et~al.}(2012){Gilmore}, {Somerville}, {Primack}, \&
  {Dom{\'\i}nguez}}]{gilmore2012a}
{Gilmore}, R.~C., {Somerville}, R.~S., {Primack}, J.~R., \& {Dom{\'\i}nguez},
  A. 2012, \mnras, 422, 3189, \dodoi{10.1111/j.1365-2966.2012.20841.x}

\bibitem[{{H.~E.~S.~S. Collaboration}(2014)}]{hess2014a}
{H.~E.~S.~S. Collaboration}. 2014, Astronomy and Astrophysics, 562, A145,
  \dodoi{10.1051/0004-6361/201322510}

\bibitem[{Hunter(2007)}]{hunter2007a}
Hunter, J.~D. 2007, Computing in Science \& Engineering, 9, 90,
  \dodoi{10.1109/MCSE.2007.55}

\bibitem[{{IceCube} {et~al.}(2018){IceCube}, {Fermi-LAT}, {MAGIC}, {AGILE},
  {ASAS-SN}, {HAWC}, {H.E.S.S.}, {INTEGRAL}, {Kanata}, {Kiso}, {Kapteyn},
  {Liverpool Telescope}, {Subaru}, {Swift/NuSTAR}, {VERITAS}, \&
  teams}]{icecube2018b}
{IceCube}, {Fermi-LAT}, {MAGIC}, {et~al.} 2018, Science,
  \dodoi{10.1126/science.aat1378}

\bibitem[{{IceCube Collaboration}(2018)}]{icecube2018a}
{IceCube Collaboration}. 2018, Science, 361, 147,
  \dodoi{10.1126/science.aat2890}

\bibitem[{{Jedamzik} \& {Saveliev}(2019)}]{jedamzik2018a}
{Jedamzik}, K., \& {Saveliev}, A. 2019, Physical Review Letters, 123, 021301,
  \dodoi{10.1103/PhysRevLett.123.021301}

\bibitem[{Jones {et~al.}(2001)Jones, Oliphant, Peterson, {et~al.}}]{scipy2001a}
Jones, E., Oliphant, T., Peterson, P., {et~al.} 2001, {SciPy}: Open source
  scientific tools for {Python}.
\newblock \url{http://www.scipy.org/}

\bibitem[{{Keivani} {et~al.}(2018){Keivani}, {Murase}, {Petropoulou}, {Fox},
  {Cenko}, {Chaty}, {Coleiro}, {DeLaunay}, {Dimitrakoudis}, \&
  {Evans}}]{keivani2018a}
{Keivani}, A., {Murase}, K., {Petropoulou}, M., {et~al.} 2018, \apj, 864, 84,
  \dodoi{10.3847/1538-4357/aad59a}

\bibitem[{{Liu} {et~al.}(2019){Liu}, {Wang}, {Xue}, {Taylor}, {Wang}, {Li}, \&
  {Yan}}]{liu2019a}
{Liu}, R.-Y., {Wang}, K., {Xue}, R., {et~al.} 2019, Physical Review D, 99,
  063008, \dodoi{10.1103/PhysRevD.99.063008}

\bibitem[{{Miniati} \& {Bell}(2011)}]{miniati2011a}
{Miniati}, F., \& {Bell}, A.~R. 2011, The Astrophysical Journal, 729, 73,
  \dodoi{10.1088/0004-637X/729/1/73}

\bibitem[{Miniati \& Elyiv(2013)}]{miniati2013a}
Miniati, F., \& Elyiv, A. 2013, The Astrophysical Journal, 770, 54,
  \dodoi{10.1088/0004-637X/770/1/54}

\bibitem[{{Murase} {et~al.}(2018){Murase}, {Oikonomou}, \&
  {Petropoulou}}]{murase2018a}
{Murase}, K., {Oikonomou}, F., \& {Petropoulou}, M. 2018, \apj, 865, 124,
  \dodoi{10.3847/1538-4357/aada00}

\bibitem[{{Neronov} \& {Semikoz}(2009)}]{neronov2009a}
{Neronov}, A., \& {Semikoz}, D.~V. 2009, Physical Review D, 80, 123012,
  \dodoi{10.1103/PhysRevD.80.123012}

\bibitem[{{Neronov} {et~al.}(2013){Neronov}, {Taylor}, {Tchernin}, \&
  {Vovk}}]{neronov2013a}
{Neronov}, A., {Taylor}, A.~M., {Tchernin}, C., \& {Vovk}, I. 2013, \aap, 554,
  A31, \dodoi{10.1051/0004-6361/201321294}

\bibitem[{{Neronov} \& {Vovk}(2010)}]{neronov2010a}
{Neronov}, A., \& {Vovk}, I. 2010, Science, 328, 73,
  \dodoi{10.1126/science.1184192}

\bibitem[{{Paiano} {et~al.}(2018){Paiano}, {Falomo}, {Treves}, \&
  {Scarpa}}]{paiano2018a}
{Paiano}, S., {Falomo}, R., {Treves}, A., \& {Scarpa}, R. 2018, The
  Astrophysical Journal Letters, 854, L32, \dodoi{10.3847/2041-8213/aaad5e}

\bibitem[{{Parma} {et~al.}(2002){Parma}, {Murgia}, {de Ruiter}, \&
  {Fanti}}]{parma2002a}
{Parma}, P., {Murgia}, M., {de Ruiter}, H.~R., \& {Fanti}, R. 2002, New
  Astronomy Reviews, 46, 313, \dodoi{10.1016/S1387-6473(01)00201-9}

\bibitem[{Plaga(1995)}]{plaga1995a}
Plaga, R. 1995, Nature, 374, 430

\bibitem[{{Saveliev} \& {Alves Batista}(2020)}]{saveliev2020a}
{Saveliev}, A., \& {Alves Batista}, R. 2020, Submitted

\bibitem[{{Sironi} \& {Giannios}(2014)}]{sironi2014a}
{Sironi}, L., \& {Giannios}, D. 2014, The Astrophysical Journal, 787, 49,
  \dodoi{10.1088/0004-637X/787/1/49}

\bibitem[{{Stecker} {et~al.}(2016){Stecker}, {Scully}, \&
  {Malkan}}]{stecker2016a}
{Stecker}, F.~W., {Scully}, S.~T., \& {Malkan}, M.~A. 2016, \apj, 827, 6,
  \dodoi{10.3847/0004-637X/827/1/6}

\bibitem[{{Takahashi} {et~al.}(2012){Takahashi}, {Mori}, {Ichiki}, \&
  {Inoue}}]{takahashi2012a}
{Takahashi}, K., {Mori}, M., {Ichiki}, K., \& {Inoue}, S. 2012, ApJ, 744, L7,
  \dodoi{10.1088/2041-8205/744/1/L7}

\bibitem[{{Takahashi} {et~al.}(2008){Takahashi}, {Murase}, {Ichiki}, {Inoue},
  \& {Nagataki}}]{takahashi2008a}
{Takahashi}, K., {Murase}, K., {Ichiki}, K., {Inoue}, S., \& {Nagataki}, S.
  2008, \apj, 687, L5, \dodoi{10.1086/593118}

\bibitem[{{Tavecchio} {et~al.}(2010){Tavecchio}, {Ghisellini}, {Foschini},
  {Bonnoli}, {Ghirlanda}, \& {Coppi}}]{tavecchio2010a}
{Tavecchio}, F., {Ghisellini}, G., {Foschini}, L., {et~al.} 2010, Monthly
  Notices of the Royal Astronomical Society, 406, L70,
  \dodoi{10.1111/j.1745-3933.2010.00884.x}

\bibitem[{{Taylor} {et~al.}(2011){Taylor}, {Vovk}, \& {Neronov}}]{taylor2011a}
{Taylor}, A.~M., {Vovk}, I., \& {Neronov}, A. 2011, \aap, 529, A144,
  \dodoi{10.1051/0004-6361/201116441}

\bibitem[{{Vallee}(1991)}]{vallee1991a}
{Vallee}, J.~P. 1991, \apss, 178, 41, \dodoi{10.1007/BF00647114}

\bibitem[{Van Der~Walt {et~al.}(2011)Van Der~Walt, Colbert, \&
  Varoquaux}]{vanderwalt2011a}
Van Der~Walt, S., Colbert, S.~C., \& Varoquaux, G. 2011, Computing in Science
  \& Engineering, 13, 22

\bibitem[{{VERITAS Collaboration}(2017)}]{veritas2017a}
{VERITAS Collaboration}. 2017, The Astrophysical Journal, 835, 288,
  \dodoi{10.3847/1538-4357/835/2/288}

\bibitem[{{Vovk} {et~al.}(2012){Vovk}, {Taylor}, {Semikoz}, \&
  {Neronov}}]{vovk2012a}
{Vovk}, I., {Taylor}, A.~M., {Semikoz}, D., \& {Neronov}, A. 2012, The
  Astrophysical Journal Letters, 747, L14, \dodoi{10.1088/2041-8205/747/1/L14}

\bibitem[{{Wang} {et~al.}(2020){Wang}, {Xi}, {Liu}, {Xue}, \&
  {Wang}}]{wang2020a}
{Wang}, Z.-R., {Xi}, S.-Q., {Liu}, R.-Y., {Xue}, R., \& {Wang}, X.-Y. 2020,
  Physical Review D, 101, 083004, \dodoi{10.1103/PhysRevD.101.083004}

\end{thebibliography}
\bibliographystyle{aasjournal}

\end{document}